\documentclass[12pt]{article}

\usepackage{amsfonts,amssymb}
\usepackage{amsmath}

\def\bbbr{{\Bbb R}}

\def\bbbc{{\Bbb C}}

\def\bbbz{{\Bbb Z}}

\def\wedgecomma{\mathop{\wedge}\limits_{'}}
\def\otimescomma{\mathop{\otimes}\limits_{'}}

\def\Ad{\mbox{Ad}\,}

\def\fr#1{{\mathfrak{#1}}}
\def\openone{\leavevmode\hbox{\small1\kern-3.3pt\normalsize1}}

\def\newpic#1{%
   \def\emline##1##2##3##4##5##6{%
      \put(##1,##2){\special{em:point #1##3}}%
      \put(##4,##5){\special{em:point #1##6}}%
      \special{em:line #1##3,#1##6}}}

\newpic{}

\begin{document}
\baselineskip=16pt


\begin{center}
{\bf REDUCTIONS AND REAL FORMS OF HAMILTONIAN SYSTEMS RELATED TO $N
$-WAVE TYPE EQUATIONS}
\end{center}

\begin{center}
{\bf V. S. Gerdjikov, G. G. Grahovski}
\end{center}

\begin{center}
{\bf \it  Institute for Nuclear Research and Nuclear Energy, Bulgarian
Academy of Sciences, 72 Tsarigradsko chaussee blvd., 1784 Sofia, Bulgaria}
\end{center}

{\bf Abstract.} Reductions of $N $-wave type equations related to
simple Lie algebras and the hierarchy of their Hamiltonian structures are
studied. The reduction group $G_R $ is realized as a subgroup of the Weyl
group of the corresponding algebra.  Some of the reduced equations are of
physical interest.

\bigskip

{\bf 1. Preliminary.}
 The analysis and the classification of all reductions for the nonlinear
evolution equations solvable by the inverse scattering method (ISM) is
interesting and still open problem.  We start with the well known form for
the $N$-wave equations \cite{ZaMa} :
\begin{eqnarray}\label{eq:1.1}
i[J,Q_t] - i[I,Q_x] + [[I,Q],[J,Q]] = 0
\end{eqnarray}
which is solvable by the ISM applied to the generalized system of
Zakharov-Shabat type:
\begin{eqnarray}\label{eq:1.2}
L(\lambda ) = id_x + [J,Q(x,t)] - \lambda J,\qquad  J \in {\frak  h}.
\end{eqnarray}
The potential matrix
\begin{eqnarray}\label{eq:1.3}
Q(x,t) = \sum_{\alpha \in \Delta _+} \left(q_{\alpha }(x,t)E_{\alpha}
 + p_{\alpha} (x,t)E_{-\alpha }\right) \in {\frak  g}\backslash
{\frak  h}
\end{eqnarray}
takes values in the simple Lie algebra ${\frak  g} $ with Cartan
subalgebra ${\frak h} $, $\Delta _+ $ is the set of positive roots of
${\frak g} $, and $E_{\alpha } $, $E_{-\alpha } $ and $H_k $ form the
Cartan-Weyl basis of ${\frak  g} $.  Indeed the $N $-wave equation
(\ref{eq:1.1}) is the compatibility condition $[L(\lambda ),M(\lambda )] =
0$, where
\begin{eqnarray}\label{eq:1.4}
M(\lambda ) = id_t + [I,Q(x,t)] - \lambda I,\qquad  I \in {\frak  h}
\end{eqnarray}

\medskip

{\bf 2.  The reduction group. }
Our basic tool is the reduction group $G_R $ introduced by A. V. Mikhailov
\cite{Mi}. $G_R$ is a finite group which preserves the Lax representation
(\ref{eq:1.3}), i.e. it ensures that the reduction constraints are
automatically compatible with the evolution. Therefore $G_R$ must have two
realizations: 1) $G_R \in \mbox{Aut}\, {\frak g} $ and 2) $G_R \in
\mbox{Conf}\, \bbbc $. To each $g_k \in G_R$ we relate a reduction
condition for the Lax pair as follows \cite{Mi}:
\begin{eqnarray}\label{eq:2.1}
C_k(L(\Gamma _k (\lambda ))) = L(\lambda ),\qquad
C_k(M(\Gamma _k (\lambda ))) = M(\lambda ),
\end{eqnarray}
where $C_k \in \mbox{Aut}\, {\frak  g}$  and $\Gamma _k(\lambda )
\in \mbox{Conf}\, \bbbc$  are the images of $g_k$ .

It is well known that $\mbox{Aut}\,{\frak   g} = {\frak  V} \otimes
\mbox{Aut}_0\,{\frak  g}$ where ${\frak  V}$ is the group of outer
automorphisms (the symmetry group of the Dynkin diagram) and
$\mbox{Aut}_0\, {\frak  g}$ is the group of inner automorphisms.  We
consider only those groups of inner automorphisms that preserve the form
of $L$ and $M$; this means that $G_R$ must preserve the Cartan subalgebra.
Then $G_R \in \fr{V}\otimes \mbox{Ad}_H \otimes W({\frak g})$ where
$\mbox{Ad}_H$ is the group of similarity transformations with elements
from the Cartan subgroup and $W({\frak g})$ is the Weyl group of ${\frak
g}$. The reductions which lead to real forms of $\fr{g} $ may be realized
by outer automorphisms of ${\frak g}$:
\begin{eqnarray}\label{eq:2.2}
C_p(X) =A_p\theta _p(X) A_p^{-1}, \qquad \Gamma _p (\lambda )=
\eta_p \lambda^*,\qquad X \in {\frak g}, \qquad p=1,2,
\end{eqnarray}
where $\theta _1(X)=X^\dag $, $\theta _2(X)=-X^* $, $A_k \in G_R$ and
$\eta_1 = 1$, $\eta _2=-1 $. Of special interest is the possibility to
embed $G_R$ in the Weyl group $W ({\frak g})$ which can be done in a
number of ways.  Therefore it is important to distinguish between the
nonequivalent reductions.  In \cite{VGN} the $\bbbz_2 $-reduced $N$-wave
systems for low-rank simple Lie algebras (${\bf A}_2 $, ${\bf C}_2 $,
${\bf G}_2 $, ${\bf A}_3 $, ${\bf B}_3 $ and ${\bf C}_3 $) are described.

\medskip

{\bf 3. Real forms as $\bbbz_2 $--reductions.}
It is well known that every real form  ${\frak g}^{\bbbr} $ can be
extracted with an involutive Cartan automorphism $\sigma $ (see e.g.
\cite{Helg}) and  $\theta (X)=X^{\dagger} $, $X\in {\frak g} $.
The automorphism that extracts a real form  of the algebra $\fr{g} $
may be viewed as a reduction (\ref{eq:2.2}) with $C_p=\sigma \theta_p$ and
$\Gamma (\lambda )= \eta _p\lambda ^* $. The compact real form
$\tilde{{\frak g}}^{\bbbr} $ of ${\frak  g} $ corresponds $\sigma =
\mbox{Id} $.  The Cartan involution splits the root system of the real
form ${\frak g}^{\bbbr} $ into two subsystems of roots: 1) compact, if
$\sigma (E_{\alpha })=E_{\alpha } $; and  2) noncompact, if $\sigma
(E_{\alpha })=-E_{\alpha }. $

Let $\pi$ be the system of simple roots of the algebra and $\pi _0 $ be
the set of compact simple roots. The Cartan involution which extracts
the noncompact real form ${\frak  g}^{\bbbr} $ from the compact one
is given by:
\begin{eqnarray}\label{eq:rf-1}
\sigma  = \exp \left(\sum_{\alpha_k \in \pi \backslash \pi _0}
{2\pi i   \over (\alpha _k,\alpha _k) } H_{\omega _k}\right) \in \Ad_H.
\end{eqnarray}
Here $H_{\omega _k}$ form basis in the Cartan subalgebra ${\frak  h}
\subset {\frak  g} $ dual to the fundamental weights $\{\omega _k\} $.
In \cite{VGN} one can find a number of examples for nontrivial reductions
of the form (\ref{eq:2.2}) where $A_k \in \Ad_H \otimes W(\fr{g}) $.

\medskip

{\bf 4. Hamiltonian structures.}
 The interpretation of the ISM as a generalized Fourier transform allows
one to prove that all $N$-wave type equations are Hamiltonian and possess a
hierarchy of Hamiltonian structures \cite{Ge} $\{H^{(k)}, \Omega ^{(k)}\}$,
$k=0, \pm 1, \pm 2, \dots $  The simplest Hamiltonian formulation of
(\ref{eq:1.1}) is given by $\{ H^{(0)},\Omega^{(0)}\}$ where $H^{(0)} =
\kappa _{0,p} (H_0 + H_{\rm int}) $ and:
\begin{eqnarray}\label{eq:3.1}
H_0 ={1 \over 2i}\int_{-\infty }^{\infty } \, dx\, \left\langle
Q,[I,Q_x] \right\rangle , \qquad
H_{\rm int} = {1\over 3} \int_{-\infty }^{\infty }\, dx \, \left\langle
[J,Q],[Q,[I,Q]] \right\rangle, \\  \label{eq:3.1.2}
\Omega^{(k)} = {i\kappa _{k,p} \over 2} \int_{-\infty }^{\infty } dx\,
\left\langle [J, \delta Q(x,t)] \wedgecomma \Lambda ^k \delta Q(x,t)
\right\rangle .
\end{eqnarray}
where $\kappa _{k,1}/\kappa^* _{k,1}=-\eta_1^k $, $\kappa _{k,2}/\kappa^*
_{k,2}=(-1)^k $ and $\langle \, \cdot\, , \cdot \, \rangle $  is the
Killing form of ${\frak  g}$ and $\Lambda  $ is the recursion operator, see
\cite{Ge}.  Then the reduction conditions (\ref{eq:2.2}) require that both
the symplectic forms and the Hamiltonians in the hierarchies are real:
$\Omega ^{(k)} = \left( \Omega ^{(k)}\right)^* $ and $H^{(k)} = \left(
H^{(k)}\right)^* $.

One of the most efficient methods to analyze the Hamiltonian structures of
the NLEE related to $L $ (\ref{eq:1.2}) is based on the classical $r
$-matrix method \cite{FaTa}. $r(\lambda ) $  is defined by:
\begin{eqnarray}\label{eq:PB}
\{ U(x,\lambda )\otimescomma U(y,\lambda ) \} = \left[
r(\lambda -\mu ), U(x,\lambda )\otimes \openone + \openone \otimes
U(y,\lambda ) \right]\delta (x-y),
\end{eqnarray}
where $U(x,\lambda )=[J,Q(x)]-\lambda J $. If the Poisson brackets between
the elements of $U(x,\lambda ) $ are the ones determined by $\Omega^{(0)}
$ then from (\ref{eq:PB}) we get $r(\lambda ) = (R_0 +R_++R_-)/\lambda  $
where:
\begin{eqnarray}\label{eq:R-m}
R_0 = \sum_{j=1}^{r} H_j \otimes H_j^\vee, \qquad
R_\pm = \sum_{\alpha \in \Delta _+} { E_{\pm\alpha } \otimes E_{\mp\alpha}
\over \langle E_\alpha ,E_{-\alpha }\rangle } ,
\end{eqnarray}
and $\langle H_j,H_k^\vee\rangle =\delta _{jk} $. Next we integrate
(\ref{eq:PB}) taking special care of the zero boundary conditions for
the potential $Q(x) $ in (\ref{eq:1.2}).  Thus for the
Poisson brackets between the matrix elements of $T(\lambda ) $
-- the scattering matrix of $L $ corresponding to $U(x,\lambda ) $ we get:
\begin{eqnarray}\label{eq:TxT}
&& \{ T(\lambda )\otimescomma T(\mu )\} = r_+ (\lambda -\mu )
T(\lambda ) \otimes T(\mu ) - T(\lambda ) \otimes T(\mu ) r_-(\lambda -\mu
), \nonumber\\
&& r_\pm(\lambda ) = {R_0  \over \lambda  } \mp i \pi \delta
(\lambda ) (R_+ - R_-).
\end{eqnarray}
Note that $R_0 $ and $\pm (R_+\pm R_-) $ are invariant with respect to
$G_R $. As a result both (\ref{eq:PB}) and (\ref{eq:TxT}) survive the
reductions for which $\Omega ^{(0)} $ remains non-degenerate, such as e.g.
reductions extracting the real forms of $\fr{g} $.

Degeneracies appear only in the case when the reduction automorphism of
the algebra maps $C(J) =-J $ and  $\Gamma (\lambda )=-\lambda  $; note
that no complex conjugation is involved here. Then both the canonical
2-form $\Omega ^{(0)}$  and the $r $-matrix become degenerate, so
the corresponding $N$-wave equations do not allow Hamiltonian formulation
with canonical Poison brackets; however they still possess a hierarchy of
Hamiltonian structures $\Omega ^{(2k+1)} $ being nondegenerate.

\medskip

{\bf 5. Example.} {\it The $N $--wave systems for ${\bf C}_2\simeq sp(4)
$ algebra.}

\begin{minipage}{3in}
\vspace*{4mm}
\special{em:linewidth 0.4pt}
\unitlength 0.75mm
\linethickness{0.4pt}
\begin{picture}(66.33,81.00)
\put(30.00,20.00){\circle{12.00}}
\put(30.00,20.00){\makebox(0,0)[cc]{10}}
\put(60.00,20.00){\circle{12.00}}
\put(60.00,20.00){\makebox(0,0)[cc]{01}}
\put(60.33,40.00){\circle{12.00}}
\put(60.33,40.00){\makebox(0,0)[cc]{11}}
\put(60.00,75.00){\circle{12.00}}
\put(60.00,75.00){\makebox(0,0)[cc]{21}}
\put(33.33,25.33){\vector(-2,-3){0.2}}
\emline{60.00}{68.67}{1}{33.33}{25.33}{2}
\put(60.00,46.33){\vector(0,-1){0.2}}
\emline{60.00}{68.67}{3}{60.00}{46.33}{4}
\put(60.00,26.33){\vector(0,-1){0.2}}
\emline{60.00}{33.67}{5}{60.00}{26.33}{6}
\put(35.33,22.67){\vector(-2,-1){0.2}}
\emline{60.00}{33.67}{7}{35.33}{22.67}{8}
\end{picture}
\end{minipage}
\begin{minipage}{3in}
{\bf Figure 1:}
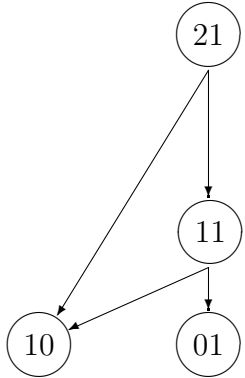
This is the wave-decay diagram for the $sp(4) $ algebra. To each positive
root of the algebra $\underline{\rm kn}\equiv k\alpha _1 + n\alpha _2 $
we put in correspondence a wave of type $\underline{\rm kn} $. If the
positive root $\underline{\rm kn} = \underline{\rm k'n'}+\underline{\rm
k''n''} $ can be represented as a sum of two other positive roots, we say
that the wave $\underline{\rm kn} $ decays into the waves $\underline{\rm
k'n'} $ and $\underline{\rm k''n''} $ as shown on the diagram to the left.

\end{minipage}

{\bf A)}
After a reduction of hermitian type (\ref{eq:2.2}) with
$A_1 =\mathrm{{diag}}\,(s_1, s_2, 1/s_2, 1/s_1)$  and $\eta_1=\pm 1 $
we obtain $a_i = \eta_1 a_i^*$, $b_i = \eta_1 b_i^* $, and
\begin{eqnarray}\label{eq:c2.hh*}
p_{10} = -\eta_1 s_1 /s_2 q_{10}^*, \qquad p_1 = -\eta_1 s_2^2q_1^*,
\qquad p_{11} =-\eta_1 s_1s_2q_{11}^*, \qquad  p_{21} =
-\eta_1s_1^2q_{21}^*.
\end{eqnarray}
The corresponding $4 $-wave system is described by the following
interaction Hamiltonian:
\begin{eqnarray}\label{eq:3.1.5}
H_{\rm int} = 4\kappa \int_{-\infty }^{\infty }\left(s_1s_2 \left(
q_{11}q_1^*q_{10}^* + \eta_1 q_{11}^*q_1q_{10} \right) - s_1^2 \left(
q_{21}q_{11}^*q_{10}^* + \eta_1 q_{21}^*q_{11}q_{10}\right) \right).
\end{eqnarray}
where $\kappa =a_1b_2-a_2b_1 $.
The case $\eta_1=1 $ and $s_1=s_2=1 $ leads to the (compact) real form
$sp(4,0) $ of ${\bf C}_2 $--algebra and it is equivalent to the $4 $-wave
interaction, see \cite{ZaMa}.

Physically we can assign to each root $\alpha  $ an wave with wave number
$k_{\alpha } $ and frequency $\omega _{\alpha } $. Each of the elementary
decays $\alpha _i=\alpha _j +\alpha _k $ is possible if $k_{\alpha _i} =
k_{\alpha _j} + k_{\alpha _k}$ and $\omega (k_{\alpha _i}) = \omega
(k_{\alpha_j}) + \omega (k_{\alpha _k})$. This can also be written using
so-called wave-decay diagrams \cite{ZaMa};  an example of such diagrams is
shown in fig.1.

{\bf B)} For $\eta=-1 $ and $s_1=s_2=1 $ if we identify $q_{10} = Q$,
$q_{11} = E_p$, $q_{21} = E_a$ and $q_{1} =E_s$, where $Q $ is the
normalized effective polarization of the medium and $E_p $, $E_s $ and
$E_a $ are the normalized pump, Stockes and anti-Stockes wave amplitudes
respectively, then we obtain the system of equations studied, e.g. in
\cite{3}. This approach allowed us to derive a new Lax pair for
(\ref{eq:3.1.5}).

\medskip

{\bf 6. Conclusions.}
 The $\bbbz_2$-reductions which act on $\lambda  $ by $\Gamma (\lambda
)=\lambda ^* $ may be viewed as Cartan involution and lead to restricting
of the system to a specific real form of the algebra ${\frak  g}$. They
preserve the canonical symplectic form $\Omega ^{(0)} $ and the classical
$r $-matrix.

The results can be extended in several directions:
i)~for more general reduction groups $G_R$; ii)~ for  NLEE with other
dispersion laws. This would allow us to study the reductions of
multicomponent NLS-type equations, two-dimensional Toda type systems etc.;
iii)~for Lax operators
with more complicated $\lambda  $-- dependence. This would allow us to
investigate also more complicated reduction groups as e.g. $\Bbb T $,
$\Bbb O $ and the possibilities to imbed them as subgroups of the Weyl
group of ${\frak  g} $.

\medskip
{\bf Acknowledgement.} The authors thank Dr. N. A. Kostov for numerous
useful discussions.

\end{document}